\documentclass[pra,onecolumn]{revtex4}
\usepackage{amsfonts}
\usepackage{amsmath}
\usepackage{amssymb}
\usepackage{graphicx}

\input{tcilatex}

\begin{document}

\title{Experimental Entanglement Distillation of Two-qubit Mixed States
Under Local filtering Operations}
\author{Zhi-Wei Wang }
\email{sdzzwzw@mail.ustc.edu.cn}
\author{Xiang-Fa Zhou }
\email{xfzhou@mail.ustc.edu.cn}
\author{Yun-Feng Huang}
\author{Yong-Sheng Zhang}
\author{Xi-Feng Ren}
\author{Guang-Can Guo }
\affiliation{Key Laboratory of Quantum Information, University of Science and Technology
of China, CAS, Hefei 230026, People's Republic of China}

\begin{abstract}
We experimentally demonstrate optimal entanglement distillation from two
forms of two-qubit mixed states under local filtering operations according
to the constructive method intruduced by F. Verstraete \emph{et al.} [Phys.
Rev. A 64, 010101(R) (2001)]. In principle, our set-up can be easily applied
to distilling entanglement from arbitrary two-qubit partially mixed states.
We also test the violation of the Clauser-Horne-Shinmony-Holt (CHSH)
inequality for the distilled state from the first form of mixed state to
show its "hidden non-locality".

PACS number(s):03.65.Ud, 03.67.Mn, 42.50.Dv, 42.65.Lm
\end{abstract}

\maketitle

Entanglement plays a key role in quantum information processing, such as
quantum teleportation \cite{1}, efficient quantum computation \cite{2} and
entangled-assisted quantum cryptography \cite{3}. In general, these
applications require maximally entangled quantum states. However, owing to
decoherence and dissipation, practical states are normally less entangled or
partially mixed. To cope with this problem, entanglement concentration is
essential \cite{zz} and various schemes have been proposed \cite{Gisin,
3H,6,4}. Experimentally Kwiat \emph{et al.} implemented entanglement
distillation from non-maximally entangled pure states and one special kind
of two-qubit mixed states using partial polarizers \cite{5}.\ After that,
entanglement concentration from maximally entangled mixed states (MEMS) has
also been realized using similar method \cite{peters}.

Generally speaking, there are two types of concentration protocols: those
involving collective operations performed on many copies of the state and
those working on individual copies. The latter case is of special interest
in practical protocols, because usually there might be technologic
difficulties to implement collective operations. For this case, although
some experiments on certain kind of two-qubit mixed states have been
experimentally demonstrated \cite{5,peters}, it is still interesting to
explore what is the best we can do to distill entanglement from arbitrary
two-qubit mixed states, only allowing local operations and classical
communication (LOCC) to be performed on each copy separately. In Ref. \cite%
{6}, Kent \emph{et al.} proved that the best state one can obtain from
general two-qubit mixed states is a Bell diagonal state. After that, F.
Verstraete \emph{et al.} constructively gave out the optimal local filtering
operations for distilling entanglement from an arbitrary two-qubit mixed
state \cite{7}. In this paper, we experimentally demonstrate optimal
entanglement distillation from two forms of two-qubit mixed states which
frequently exist in the real world. The optimal local filtering operations
are explicitly calculated according to the method provided in Ref. \cite{7}.
And in principle, our set-up can be applied to entanglement distillation of
arbitrary two-qubit mixed states.

For general two-qubit mixed states, consider two separated parts, Alice and
Bob, who each control one subsystem and are only allowed to carry out local
operations and classical communications. Specifically, Alice and Bob are
only permitted to perform local unitary transformations and local
filterings. Then what we are concerned about is whether they can increase
the entanglement of the system and to what extent they can do that under
LOCC. This problem is very interesting because any real world quantum
communication channel is imperfect and generally, Alice and Bob are not able
to share perfect maximally entangled states directly.

Theoretically, by introducing the real and linear parametrization, a
two-qubit density matrix can be represented as $\rho =\frac{1}{4}%
\sum_{i,j=0}^{3}R_{ij}\sigma _{i}\otimes \sigma _{j}$, where $\sigma _{0}$
is the $2\times 2$ identity and $\sigma _{1},\sigma _{2},\sigma _{3}$ are
usual Pauli matrices \cite{7}. LOCC of the type $\rho ^{^{\prime }}\sim
(A\otimes B)\rho (A\otimes B)^{\dag }$ corresponds to left and right
multiplication by a proper orthochronous Lorentz transformation (POLT) ,
followed by normalization in this $R$-picture. The concentration protocol,
that is optimal in the sense that it produces a state of maximum
entanglement of formation (EoF) \cite{6}, has been obtained in \cite{7} for
two cases. If $R=[R_{ij}]$ ($R_{ij}=Tr(\rho (\sigma _{i}\otimes \sigma _{j}))
$) is diagonalizable by POLT, a Bell diagonal mixed state can be extracted
from the input state with a finite probability, which has the maximum EoF
and the maximum possible violation of the Clauser-Horne-Shimony-Holt (CHSH)\
version of inequality \cite{13,wolf}. When $R$ is not diagonalizable by
POLT, the initial state can be quasi-distillable. In the extreme case, the
input state can be asymptotically transformed into a Bell diagonal state
with lower rank, while the success probability becomes infinitesimally close
to zero. This method can also be used for entanglement concentration from
MEMS since the method in \cite{peters} is a particular case of it.\textbf{\
\ \ \ \ \ \ \ \ \ \ \ \ \ \ \ \ \ \ \ \ \ \ \ \ \ \ \ \ }

Experimentally, to demonstrate the optimal entanglement distillation
protocol, we concentrate on the mixed states that can be diagonalized by
POLT. We specially devise two forms of two-qubit mixed states which can be
easily prepared and distilled under fewer local operations. The experiment
set-up to investigate entanglement distillation is shown in Fig. $1$. A $0.59
$ $mm$ thick $\beta $-barium borate (BBO) crystal arranged in the Kwiat type
configuration \cite{8} is pumped by a $351.1$ $nm$ laser beam produced by an
Ar$^{\text{+}}$ laser. In the spontaneous parametric down-conversion (SPDC)
process, a nonmaximally entangled state $a|HH\rangle +b|VV\rangle $ ( $H$
and $V$ represent horizontal and vertical polarization of the photons
respectively) is produced. In the experiment, for simplicity, we adjust the
half wave plate (HWP) and tiltable\ quarter wave plate (QWP) in the pump
path to choose the real numbers $a$ and $b$ \cite{8}. The experiment can
also be applied to the situation where $a$ and $b$ are complex numbers.\
After the BBO crystal, the photon pairs pass through two phase-damping
channels in a particular basis. In our scheme, each phase-damping channel
consists of one quartz plate with its\ optical axis rotated by a certain
angular to the vertical axis. For the first form of mixed state, we select
two same phase-damping channels in $\{H+V,H-V\}$ basis with the
corresponding super-operators $\{\sqrt{1-p}I,\sqrt{p}\sigma
_{x}\}_{A}\otimes \{\sqrt{1-p}I,\sqrt{p}\sigma _{x}\}_{B}$, where $p$ is
connected with the thickness of the quartz and the bandwidth of the
interference filter \cite{9}. Then the mixed state has the form 
\begin{equation}
\rho _{I}=\left( 
\begin{array}{cccc}
(p-1)^{2}+b^{2}(2p-1) & 0 & 0 & ab((p-1)^{2}+p^{2}) \\ 
0 & p-p^{2} & 2abp(1-p) & 0 \\ 
0 & 2abp(1-p) & p-p^{2} & 0 \\ 
ab((p-1)^{2}+p^{2}) & 0 & 0 & p^{2}-b^{2}(2p-1)%
\end{array}%
\right) \text{,}
\end{equation}%
In the experiment, we choose $a=0.23$, $b=0.97$, and $p=0.013$ (a $\sim 1.5$%
-mm-thick quartz plate).

According to Ref. \cite{7}, the local operations for Alice and Bob are$%
\left( 
\begin{array}{cc}
1 & 0 \\ 
0 & 0.49%
\end{array}%
\right) $ and $\left( 
\begin{array}{cc}
0 & 1 \\ 
1 & 0%
\end{array}%
\right) \left( 
\begin{array}{cc}
1 & 0 \\ 
0 & 0.49%
\end{array}%
\right) $ respectively. For Alice, the filtering operation can be realized
by inserting into her path microscope slides tilted about the vertical axis $%
73^{\circ }$. This configuration has a measured transmission of $92\%$ for $H
$ polarization and $22\%$ for $V$ polarization. As for Bob, there are one
local filtering and one single-qubit unitary operation. He just needs to
place a half wave plate (HWP) after the microscope slides with the angle
between the\ optical axis and the\ vertical axis set to $45^{\circ }$. We
can also realize the local filtering operation using a Mach-Zehnder
interferometer \cite{zhang}, but it requires too much about its stability
and visibility.

To obtain the density matrices of the distilled states, we use the
technology of maximum likelihood tomography \cite{8,10}. By means of a
quarter wave plate (QWP), HWP and polarizing beam splitter (PBS) in each
arm, the polarization of each photon can be analyzed in an arbitrary basis.
Then by combinations of 16 single-photon projections $|H\rangle $, $%
|V\rangle $, $|H\rangle +|V\rangle $, $|H\rangle +i|V\rangle $ on each of
the two-photon, we may derive the density matrices describing the states of
the photon pairs (all the experimental density matrices in this paper can be
found in \cite{buchong}). Fig. $2$ shows the density matrices before and
after entanglement distillation.

Here we use concurrence to characterize the entanglement between the two
subsystems since the EoF of a two-qubit system is a convex monotonically
increasing function of the concurrence \cite{11}. The concurrence before and
after distillation are $0.248\pm 0.021\ $and\ $0.672\pm 0.044$\
respectively. The distilled state has the fidelity of \cite{12} $82\%$ with
the theoretical Bell diagonal state.\ From the data we can see that by means
of the local filtering operations we can effectively increase the
entanglement between the two subsystems. The errors mainly stem from the
imperfect preparation of the initial mixed state\ (the fidelity is $94\%$)\
and in the case of strong filtering, small changes in the initial state may
have a large impact on the final state.

We also show the \textquotedblleft hidden non-locality\textquotedblright\ of
the distilled state by its violation CHSH\ version of inequality. In the
CHSH inequality, the proposed value $|S|$,$\ $a combination of four
polarization correlation probabilities, should not be more than 2 for local
hidden variables theory. If $|S|>2$,$\ $we can only use quantum mechanics to
explain the correlations. Experimentally, the set-up for tomographic
reconstruction can also be used for the measurement of the CHSH inequality.
The analysis settings of QWPs and HWPs are determined by the tomographically
measured density matrix of each state using the method introduced in Ref. 
\cite{14}. Following the normalization procedure \cite{15}, we obtain the
value $S_{before}=1.853\pm 0.011$\ and $S_{after}=2.175\pm 0.024$,
corresponding to the values of $S$ before and after the entanglement
distillation. The ideal value of $S_{after}$\ is $2.192$. The experiment
result accords well with the theory. Thus we obtain the violation of about $%
7\sigma $.\ In the process of entanglement distillation, Alice and Bob only
employ LOCC operations. As pointed by Kwiat \cite{5}, the non-locality
demonstrations rely on\ \textquotedblleft conditional
probabilities\textquotedblright : we use LOCC operations to select a
subensemble with a small probability which can demonstrate non-local
correlations. This idea is similar to entanglement distillation: we can just
increase the entanglement of subensemble and can not do that for the whole
ensemble \cite{16}.\ 

Next consider distilling the entanglement from the other form of two-qubit
mixed state. This time we let the nonmaximally entangled states pass through
two same phase-damping channels in $\{H,V\}\ $basis.\ The corresponding
super-operators are $\{\sqrt{1-p}I,\sqrt{p}\sigma _{z}\}_{A}\otimes \{\sqrt{%
1-p}I,\sqrt{p}\sigma _{z}\}_{B}$.\ Each channel is a $\sim 3$ $mm$\ quartz
which has a polarization-dependent optical path length difference about $%
40\lambda $ ($\lambda =702.2$ $nm$). After the decoherence, the state has
the form%
\begin{equation}
\rho _{II}=\left( 
\begin{array}{cccc}
a^{2} & 0 & 0 & ab(-1+2p)^{2} \\ 
0 & 0 & 0 & 0 \\ 
0 & 0 & 0 & 0 \\ 
ab(-1+2p)^{2} & 0 & 0 & b^{2}%
\end{array}%
\right) \text{.}
\end{equation}%
For this kind of mixed states, the only local operation is just a unilateral
local filtering $\left( 
\begin{array}{cc}
b & 0 \\ 
0 & a%
\end{array}%
\right) $\ for Bob. We consider two cases ($a=0.44,$ $0.52$ and $p=0.063$).
The experiment results are shown in Fig. $3$. Under this local filtering
operation, the entanglement of the final states can be increased. For $%
a=0.44 $, the concurrence increases from $0.552\pm 0.017$ to $0.641\pm 0.022$%
; the fidelities with the corresponding theoretical states are $98\%$ and $%
96\%$ respectively. For $a=0.52$, the concurrence increases from $0.569\pm
0.017$ to $0.666\pm 0.021$; the fidelities with the corresponding
theoretical states are $97\%$ and $97\%$ respectively.

In conclusion, we have experimentally demonstrated optimal local filtering
operations for two general forms of two-qubit mixed states that can be
diagonalized by POLT. For the first form of mixed state, the local
operations involve bilateral filtering; while for the second one, only
unilateral filtering is needed. In fact, we can generalize this method to
arbitrary two-qubit partially mixed states, because any nontrivial LOCC
operations have the form \cite{6,zhang} $\gamma U_{A}\left( 
\begin{array}{cc}
1 & 0 \\ 
0 & \alpha _{A}%
\end{array}%
\right) U_{A}^{\prime }\otimes U_{B}\left( 
\begin{array}{cc}
1 & 0 \\ 
0 & \alpha _{B}%
\end{array}%
\right) U_{B}^{\prime }$, where\ $U_{A(B)}$ and $U_{A(B)}^{\prime }$\ denote
local unitary operations for Alice (Bob) and $\gamma $ is a scale factor in
the range $0\leq \gamma \leq 1$ and$\ 0\leq \alpha _{A(B)}$\ $\leq 1$.\ The
main experiment difficulity lies in the preparation of arbitrary two-qubit
mixed states.

On the other hand, the imperfect preparation of initial mixed states is also
a main reason for the deviations of our results from theoretical
calculation. However, in practical schemes, one can directly choose the
states derived from the tomographic measurment of the input as the initial
mixed states and implement corresponding local filtering operations on them.
Thus the experiment results will approach the theory with better accuracy.
After the local filtering operations, Bell diagonal states can be obtained
from the input states\ which have the maximum concurrence and the maximum
possible violation of the CHSH inequality. We experimentally obtain the
violation of the CHSH inequality about $7\sigma $\ for the distilled state
from the first form of mixed state, which verifies\ \textquotedblleft hidden
non-locality\textquotedblright\ for this form of mixed state in the process
of entanglement distillation. Recently it has been found that Bell diagonal
states can also be used to implement the generalized tomographic quantum key
distribution protocol \cite{Dagomir}. The method used in this experiment is
very simple and universal since it just needs LOCC operations and it can
effectively increase the entanglement of arbitrary two-qubit partially mixed
states. We believe it will be helpful in the exploration of various quantum
information processing.

The authors would like to thank Fang-Wen Sun and Guo-Yong Xiang for helpful
discussions. This work was funded by the National Fundamental Research
Program (2001CB309300), National Natural Science Foundation of China
(10304017, 10404027, 60121503), program for New Century Excellent Talents in
University, the Innovation Funds from Chinese Academy of Sciences.\ \ \ \ \
\ \ \ \ \ \ \ \ \ \ \ \ \ \ \ \ \ \ \ \ \ \ \ \ \ \ \ \ \


\begin{thebibliography}{99}
\bibitem{12} Here the fidelity\ between two states $\rho _{in}$ and $\rho
_{out}$ is defined as $|Tr(\sqrt{\sqrt{\rho _{in}}\rho _{out}\sqrt{\rho _{in}%
}})|^{2}$.
\end{thebibliography}
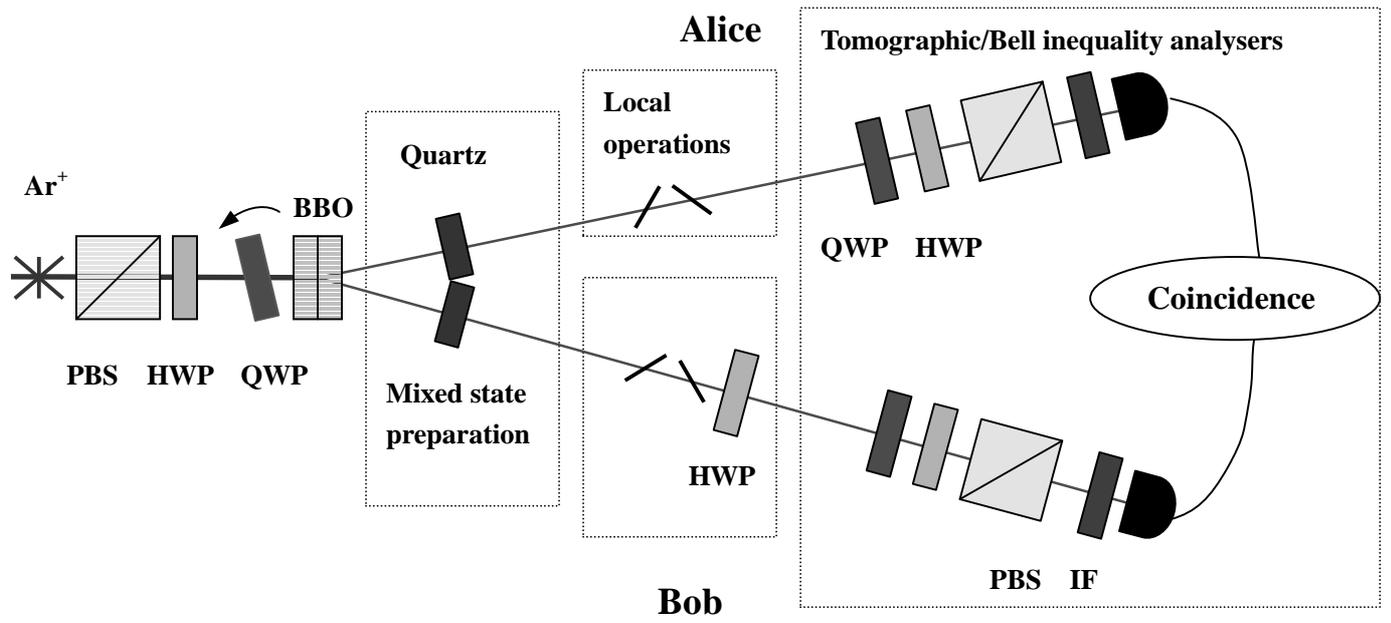
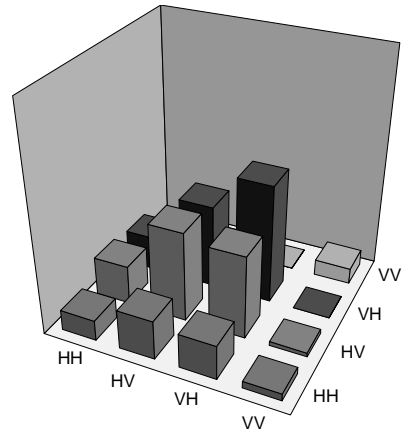
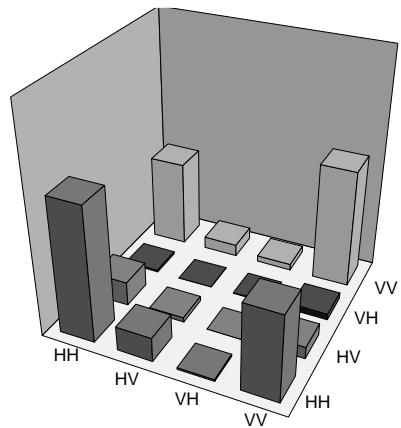
\end{document}